\documentclass{aa}
\usepackage{psfig}
\usepackage{natbib}
\bibpunct{(}{)}{;}{a}{}{,}

\def\ecs{erg~cm$^{-2}$s$^{-1}$}
\def\lum{erg~s$^{-1}$}
\def\xr{$\chi^2_\mathrm{r}\,$}

\begin{document}
\bibliographystyle{aa}

\title{Burst-properties as a function of mass accretion rate in GX~3+1}

\author{P.R. den Hartog\inst{1,2}  
  \and J.J.M in 't Zand\inst{1,2}  
  \and E. Kuulkers\inst{1,2}\thanks{  
    \emph{Present address:} ESA-ESTEC, SCI-SDG, Keplerlaan~1, 2201~AZ,  
    Noordwijk, The Netherlands} 
  \and R. Cornelisse\inst{1,2} 
  \and J. Heise\inst{2,1}
  \and A. Bazzano\inst{3}
  \and M.~Cocchi\inst{3}
  \and L. Natalucci\inst{3}
  \and P. Ubertini\inst{3}}
\offprints{P.R. den Hartog}
\mail{P.R.den.Hartog@sron.nl}

\institute{Astronomical Institute, Utrecht University,
              P.O.Box 80000, 3508 TA Utrecht, The Netherlands
        \and SRON National Institute for Space Research, Sorbonnelaan 2,
              3584 CA Utrecht, The Netherlands
        \and Istituto Astrofisica Spaziale e Fisica Cosmica, CNR,
              Via Fosso del Cavaliere 100, I-00133 Roma, Italy}

\date{\today / Accepted date}

\abstract{ \object{GX~3+1} is a low-mass X-ray binary that is persistently
  bright since its discovery in 1964. It was found to be an X-ray burster
  twenty years ago proving that the compact object in this system is a neutron
  star. The burst rate is so low that only 18 bursts were reported prior to
  1996.  The Wide Field Cameras on {\em BeppoSAX} have, through a dedicated
  monitoring program on the Galactic center region, increased the number of
  X-ray bursts from GX~3+1 by 61. Since GX~3+1 exhibits a slow (order of
  years) modulation in the persistent flux of about 50\%, these observations
  opens up the unique possibility to study burst properties as a function of
  mass accretion rate for very low burst rates.  This is the first time that
  bursts are detected from GX~3+1 in the high state. From the analysis we
  learn that all bursts are short with e-folding decay times smaller than 10
  s. Therefore, all bursts are due to unstable helium burning. Furthermore,
  the burst rate drops sixfold in a fairly narrow range of 2--20 keV flux; we
  discuss possible origins for this.

}

\maketitle

\section{Introduction}
\label{sec:intro}

GX~3+1 is one of the first discovered cosmic X-ray sources. It was detected
during an {\em Aerobee}-rocket flight on June 16, 1964 \citep{Bowyer65}.  Ever
since, GX~3+1 is one of the brightest persistent sources with an average
$2-10\,\rm{keV}$ flux half that of the Crab.  Its X-ray intensity shows a mild
variability on an hourly time scale and possibly a sinusoidal-like variation
on a time scale of years \citep{Makishima83}.

Although an optical counterpart has not yet been identified
\citep[e.g.,][]{Naylor91}, GX~3+1 is presumably a low mass X-ray binary (LMXB)
where a neutron star (NS) is accompanied by a low-mass star of spectral type A
or later. In such a system \citep[for a review, see][]{White95} the companion
overflows its Roche lobe and dumps matter onto the NS through an accretion
disc. The accumulation of matter on the NS surface leads to pressures and
temperatures that are high enough to initiate nuclear fusion processes of the
accreted hydrogen and helium to higher-Z nuclei \citep{Hansen75}. The
circumstances of this process are such that the fusion may be unstable; only
at accretion rates in excess of the Eddington limit is the burning thought to
be completely stable (for solar abundances of the accreted matter; Fujimoto et
al. 1981).  Unstable fusion occurs through thermonuclear flashes which may be
observable in the form of X-ray bursts \citep{Woosley76,Maraschi77}.  Such
so-called type-I x-ray bursts are characterized by a fast rise (one to ten
seconds), exponential-like decay (e-folding decay times between a few seconds
and tens of minutes), a black body spectrum of $kT\sim$1--3~keV, and cooling
during the decay \citep[for a review, see][]{Lewin93}.  For the flash to be
observable, helium is essential as a fuel ingredient \citep{Joss77}. The
amount of hydrogen involved in the burning layer determines the longevity of
the explosion \citep[e.g.,][]{Taam79}.  Currently, about half of all LMXBs
have been seen to burst \citep[e.g.,][]{Zand01}; GX~3+1 is one of them.

GX 3+1 was first seen to burst in 1983 (Makishima et al. 1983), about eight
years after the discovery of the first X-ray burster \citep{Grindlay75},
testifying to an infrequent bursting behavior. Up to now 20 bursts have been
reported in the literature from GX~3+1: 15 with {\em Hakucho}
\citep{Makishima83}, 2 with {\em Ginga} \citep{Asai93}, 1 with {\em Granat}
\citep{Pavlinsky94} and 2 with the {\em Rossi X-ray Timing Explorer} (RXTE)
\citep{Kuulkers00,Kuulkers02}.  Nineteen of these bursts are short with decay
times less than 10 seconds.  The remaining burst, seen with the All Sky
Monitor (ASM) on RXTE, is a superburst with a decay time of $\sim$1.6 hours
\citep{Kuulkers02}.  One of the short bursts was seen to exhibit a quick
(i.e., less than 2 seconds) radius expansion (factor of two) phase, indicating
that the burst luminosity was near or at the Eddington luminosity, causing the
NS atmosphere to expand due to radiation pressure. This
implies a distance to the source of 4.5 kpc, for a hydrogen-rich NS atmosphere
\citep{Kuulkers00}.

In a simple picture a larger accretion rate would imply that ignition
conditions are reached in quicker succession and burst rates should
correspondingly be higher, under the assumption that the amount of
fuel burned per burst is similar. Remarkably, the opposite is usually
observed. In the majority of the roughly ten cases where this issue
could be investigated (including GX~3+1) the burst frequency was found
to be {\it smaller} at higher accretion rates, assuming that higher
fluxes reflect higher accretion rates.  Only in one case is there
clear evidence for a positive correlation: GS~1826-24
\citep{Cocchi00}. The anti-correlation between flux and burst
frequency prompted the fuel 'leakage' hypothesis: fuel is burnt in a
non-observable way with a rate that increases with accretion rate
\citep{Paradijs88}. Recently, \citet{Bildsten00} proposed as an
alternative that the explosion may be occurring on only part of the NS
surface, and that the expected proportionality for the burst frequency
is with the accretion rate {\em per unit area} instead of the global
accretion rate \citep[see also][]{Marshall82}.

We here discuss observations of GX~3+1 that were obtained with the Wide Field
Cameras on {\em BeppoSAX} between 1996 and 2002. This data set is unique
because of its sheer volume, giving rise to the detection of 61 bursts, which
allows for a more detailed study of the infrequent bursting behavior of GX~3+1
than was hitherto possible.  The merits of such a study are increased by the
slow but large change in brightness over the course of these observations,
enabling an analysis of the dependency of burst properties, such as recurrence
times, on flux and possibly accretion rate.

\section{Observations}
\label{sec:obs}

\subsection{{\em BeppoSAX} observations}
\label{sec:obs_sax}

The Italian-Dutch X-ray satellite {\em BeppoSAX} \citep{Boella97a}, launched
on April 30, 1996, and switched off exactly six years later, carried two Wide
Field Cameras \citep[WFCs;][]{Jager97}.  They were active between 2 and 28 keV
and had field of views (FOVs) of $40\degr \times 40\degr$ with an angular
resolution of $5'$, and a source location accuracy of $0.7'$ (99\%
confidence).  One of the core programs of {\em BeppoSAX} was a Galactic center
monitoring program with the WFCs. The observations were made during two
visibility windows per year, from mid February to mid April and from mid
August to mid October (see Table \ref{tab:obs} and Fig. \ref{fig:asm}). Since
GX~3+1 is located close to the Galactic center, it was usually near the center
of the FOV, so that the sensitivity was optimum. The complete dataset consists
of 12 Galactic center campaigns, including serendipitous secondary observations
near the Galactic center, with a total net exposure time of 6.2 Ms. In this
dataset 61 bursts were identified with GX~3+1.  The bursts were found through
analyzing the full bandpass (2--28 keV) light curves of GX~3+1 at a time
resolution of 5~seconds. The peak flux detection threshold for bursts that
have an e-folding decay time of at least 3~seconds is then expected to be
$1\,\rm{cts\,cm^{-2}\,s^{-1}}$, or roughly half the flux of the Crab.

\begin{table*}
\caption[]{Summary of the twelve Galactic center observation campaigns of 
  the WFCs. For every campaign it shows the campaign number, year, begin and
  end time ($t_{\mathrm{begin}}$ and $t_{\mathrm{end}}$), the number of 
  observations, the net exposure time ($t_{\mathrm{exp}}$), the number of 
  bursts seen from GX~3+1 during that campaign, the derived burst rate and 
  the average persistent ASM intensity ($\overline{F_{\rm{pers}}}$).}

\begin{tabular}{r@{}l l l@{}l l@{}l r@{}l r@{}l r@{}l r@{}l@{} l}

\hline 

\vspace{-3mm}\\

\multicolumn{2}{l}{Camp.} & 
year & 
\multicolumn{2}{l}{$t_{\mathrm{begin}}$} &
\multicolumn{2}{l}{$t_{\mathrm{end}}$} & 
\multicolumn{2}{r}{\# obs.} & 
\multicolumn{2}{r}{$t_{\mathrm{exp}}\,(\rm{ks})$} & 
\multicolumn{2}{r}{\# bursts} & 
\multicolumn{2}{r}{burst rate ($\rm{Ms^{-1}}$)} & 
$\overline{F_{\rm{pers}}}\,(\mathrm{cts\,s^{-1}})$\\

\hline

\vspace{-3mm}\\

&1 & 1996 & Aug.&15 & Oct.&29 & 6&7 & 101&7 & &3 & 2.9&$^{+2.9}_{-1.6}$ & $25.13\pm0.06$\\
&2 & 1997 & Mar.&02 & Apr.&26 & 2&1 & 65&4 & &3 & 4.6&$^{+4.4}_{-2.5}$ & $25.23\pm0.07$\\
&3 & 1997 & Sep.&06 & Oct.&12 & 1&3 & 30&2 & &0 & $<$3.8& & $22.22\pm0.09$\\
&4 & 1998 & Feb.&11 & Apr.&11 & 1&7 & 55&1 & 1&2 & 21.8&$^{+8.2}_{-6.2}$ & $18.25\pm0.07$\\
&5 & 1998 & Aug.&22 & Oct.&23 & 1&0 & 41&0 & &8 & 19.5&$^{+9.6}_{-6.7}$ & $14.98\pm0.07$\\
&6 & 1999 & Feb.&14 & Apr.&11 & 1&4 & 47&0 & &8 & 17.2&$^{+8.3}_{-5.8}$ & $14.92\pm0.08$\\
&7 & 1999 & Aug.&24 & Oct.&17 & 2&4 & 80&1 & 1&9 & 23.7&$^{+6.7}_{-5.3}$ & $15.50\pm0.08$\\
&8 & 2000 & Feb.&18 & Apr.&07 & 2&1 & 63&3 & &3 & 4.7&$^{+4.6}_{-2.5}$ & $20.47\pm0.09$\\
&9 & 2000 & Aug.&22 & Oct.&16 & 2&9 & 76&7 & &2 & 2.6&$^{+3.4}_{-2.5}$ & $25.82\pm0.09$\\
1&0 & 2001 & Feb.&14 & Apr.&23 & &5 & 21&5 & &1 & 4.7&$^{+10.6}_{-3.9}$ & $25.64\pm0.07$\\
1&1 & 2001 & Sep.&04 & Sep.&30 & &7 & 28&4 & &2 & 7.0&$^{+8.9}_{-4.9}$ & $25.57\pm0.08$\\
1&2 & 2002 & Mar.&05 & Apr.&15 & &5 & 9&1 & &0 & $<$12.5& & $25.33\pm0.14$\\
\hline
\end{tabular}

\label{tab:obs}
\end{table*}

\subsection{ASM observations}
\label{sec:obs_asm}

\begin{figure}
\psfig{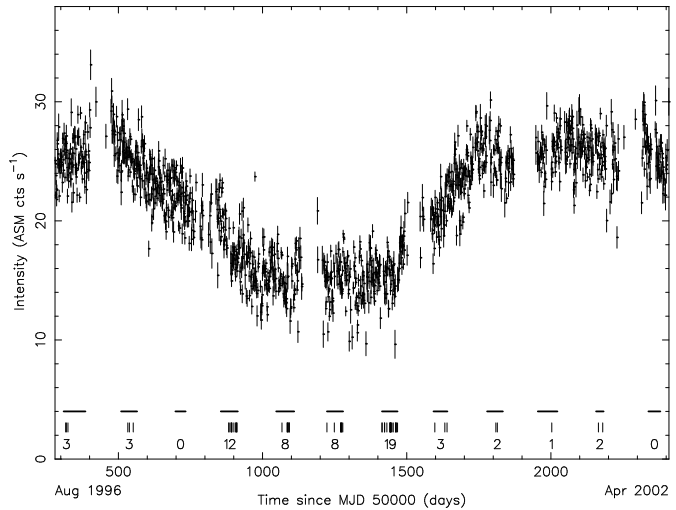}
\caption{ASM 1.5--12 keV light curve (one-day averages) from August 1996 to
  April 2002. For clarity, data points with error bars larger than
  1.25~$\rm{cts\,s^{-1}}$ were not plotted (i.e., 16\% of all points). Below
  the light curve the time spans of the twelve {\em BeppoSAX} Galactic center
  campaigns are shown with horizontal bars.  Below the time spans the burst
  times are indicated with vertical lines, and we indicate the number of
  bursts per campaign.
\label{fig:asm}}
\end{figure}

The All Sky Monitor on board RXTE \citep{Levine96} monitored GX~3+1 over the
six years covering the WFC era (and beyond). Fig. \ref{fig:asm} shows the 1.5
to 12 keV light curve.  Plotted are one-day averages of the 90 second dwells
taken by the ASM. It shows that the intensity of GX~3+1 slowly changed by a
factor of two.  At first the source was in a 'high state' with a flux of about
$0.4$ times that of the Crab ($75\,\rm{cts\,s^{-1}}$).  Then the intensity
dropped over the course of about 20 months until it reached a 'low state' with
a flux that is 50\% lower.  GX~3+1 stayed in this low state for about 19
months after which the intensity slowly increased until it reached the high
state after 10 months again. The reason for using the ASM data set is that it
provides a more uniform coverage than the WFC data set. Thus, one can get a
better idea how the source behaved between bursts when the recurrence time is
longer than the typical WFC observation.

\section{Burst analysis}
\label{sec:burst}

A time-resolved spectral analysis was performed of all 61 bursts. The time
resolution is limited by the statistical quality of the data, which is
determined by the burst flux, duration and the off-axis angle. We required a
resolution so that the 1 sigma errors in the spectral parameters remained
within 20\%.

The 2--28~keV light curves, at a resolution of one second, served as a guide
for setting up the time intervals for which burst spectra were extracted.
The first interval was set to start when the photon flux showed a
significant trend towards a clear peak.  In this regard it is useful to
note that all 61 bursts showed onsets that lasted less than three seconds. The
end of the burst was defined as the instance when an exponential fit function
to the decay dropped to the average persistent level within one sigma. We also
included two time intervals covering the persistent emission, namely a 10--15
seconds interval just before the burst onset and a 80--90 seconds interval
from the end of the burst. In total three to five time bins were defined for
each burst.  For each time interval, dead-time corrected background-free
spectra were extracted in the 2--28 keV band. During the analysis we ignored
channels 1, 2 and 31, and included a systematic error of 1.5\% per channel, to
accommodate calibration uncertainties.

\begin{figure}
\psfig{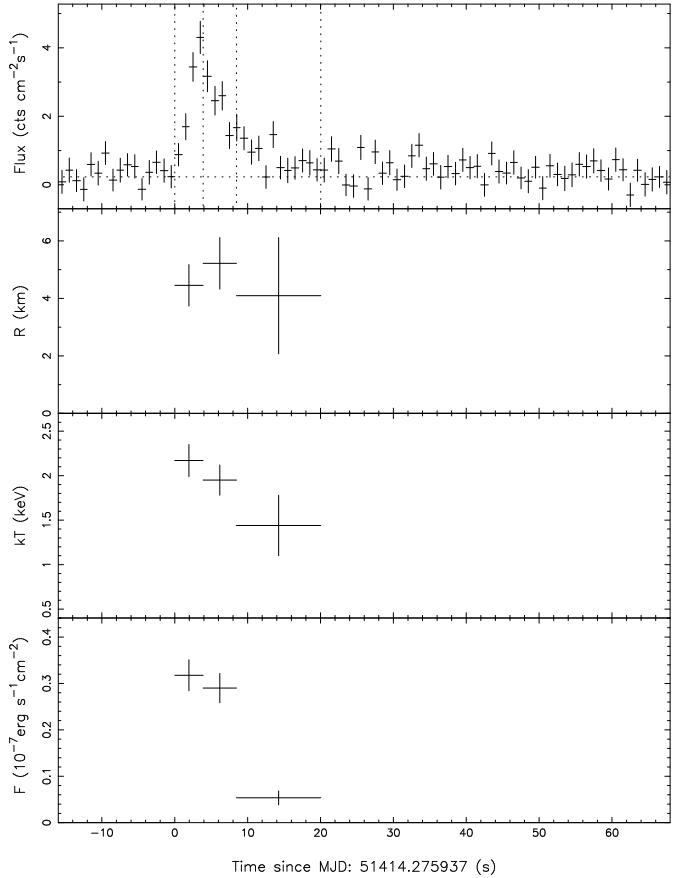}
\caption{Example of the results of the spectral analysis of a burst. In the
  top panel the 2--28 keV light curve of the burst is shown. The horizontal
  dotted line is the determined persistent flux. The vertical dotted lines
  show the bin boundaries. The second panel shows the radii of the emitting
  region on the NS (assuming isotropic emission and a distance of 4.5 kpc).
  The radii of the three bins are constant over time. The third panel shows
  the black body temperature of the NS. In the bottom panel the average
  bolometric black body flux is shown.
\label{fig:totplot}}
\end{figure}

For each burst, we assumed that the spectrum of the persistent emission was
constant and modeled it with thermal bremsstrahlung.  This is a simplification
with respect to the more sophisticated models discussed in section
\ref{sec:pers}, but is allowed because 1) the statistical quality of the short
(115 sec) segments is insufficient to discriminate between models, and 2) we
are not interested in the details of this spectrum at this point but merely in
(implicitly) subtracting the persistent emission. The spectrum of the burst
emission was modeled by a single-temperature black body, whereby the
temperature and normalization were left free over all burst time intervals. On
all spectral models, a fixed absorption was applied, following
\citet{Morrison83}, with $N_{\rm H}=1.59\times10^{22}$~cm$^{-2}$
\citep{Oosterbroek01}.  This model fits the data satisfactorily (\xr ranges up
to 1.5 with at least 80 dof for at least 3 spectra). For each burst interval
the fit provides an estimate for the radius ($R$) of the emission area (i.e.,
of a sphere and assuming isotropical emission), the black body temperature
($kT$) and the predicted unabsorbed bolometric flux ($F$). An example of the
results of the spectral analysis is shown in Fig.  \ref{fig:totplot}. Since
the time bins are broad, the parameters are not very time sensitive.  For
instance, there is not a high enough temporal resolution to detect a radius
expansion such as seen by \citet{Kuulkers00}, which lasted less than two
seconds. Also the peak bolometric flux is lower than the true peak flux for
the same reason.  Therefore, in the remainder we will consider only the
observed peak photon flux ($F_{\rm peak}$), since in this paper we only
compare bursts seen by the WFC. The total fluence ($E_\mathrm{b}$) is
determined by integrating the mean observed fluxes over the time bins.

\begin{figure}
\psfig{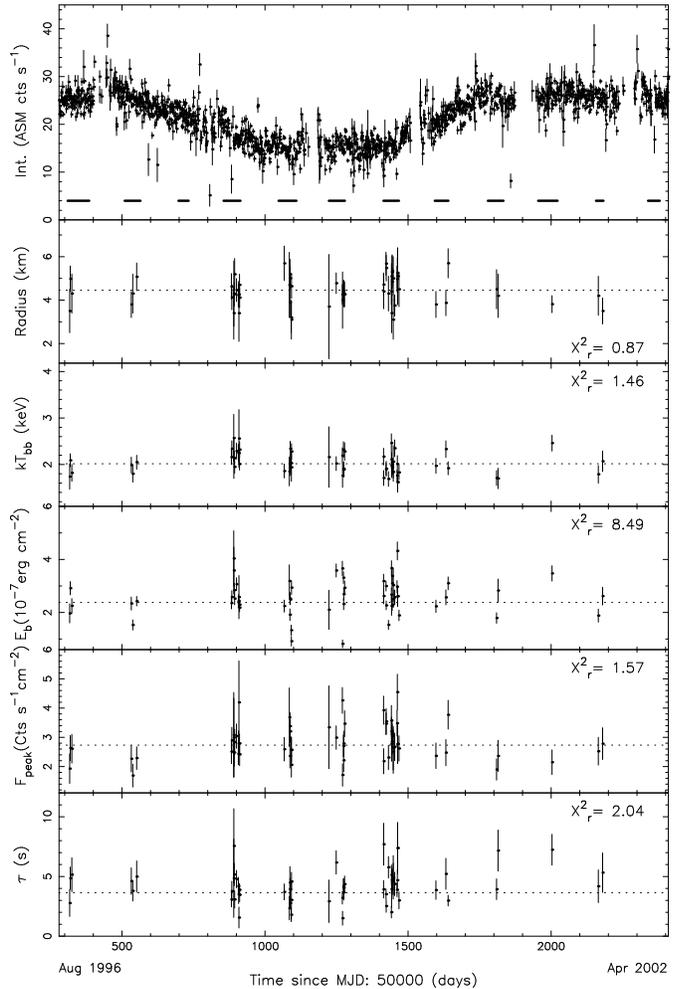}
\caption{In the top panel the ASM 1--12 keV light curve is shown together with
  the WFC coverage as horizontal bars. In the lower panels the burst
  properties versus time. From 2nd to lowest panel: radius of the emission
  area (assuming 4.5 kpc distance), black body temperature, burst fluence,
  peak flux and decay time. The dotted lines are the weighted means of the
  properties. The radii per burst are the weighted means of all time bins,
  wile the temperature is derived from the first time bin since the burst
  onset.
\label{fig:overview}}
\end{figure}

\begin{figure}
\psfig{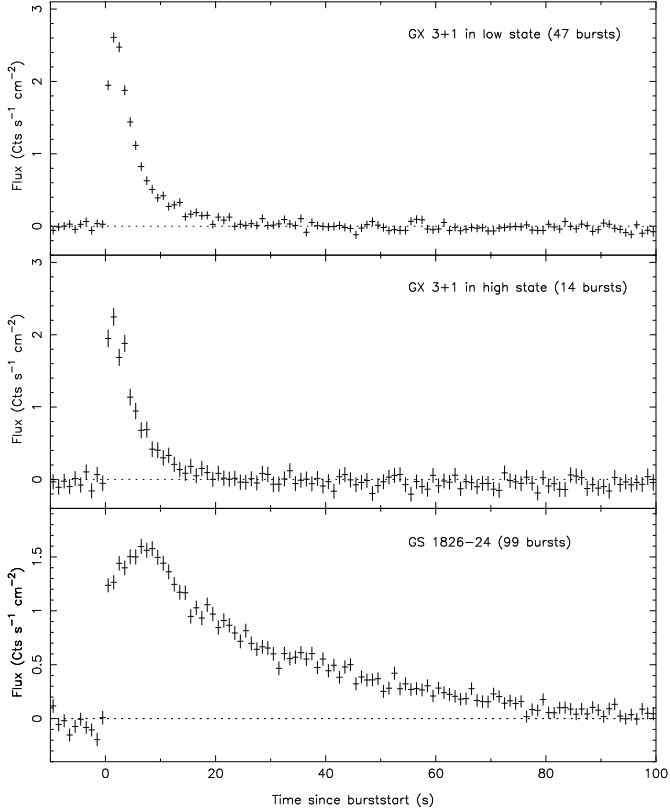}
\caption{Comparison of average 2--28 keV burst profiles. From top to
    bottom, the average profiles of GX~3+1 in the low state, GX~3+1 in the
    high state and \mbox{GS1826-24} are given. The average decay times are
    $4.23 \pm 0.13$ (\xr~=~1.15), $4.97 \pm 0.28$ (\xr~=~0.61) and $29.21 \pm
    0.97$ (\xr~=~1.13) seconds, respectively.
\label{fig:lc_sum}}
\end{figure}

Fig. \ref{fig:overview} shows the burst parameters for all 61 bursts plotted
versus time. To search for variability in each parameter, \xr was determined
with respect to the weighted mean.  This shows that the emission area is
consistent with being constant (\xr $=$ 0.87).  The weighted mean of its
spherical radius is $4.46\ \pm 0.08\,\rm{km}$ for a distance of 4.5 kpc, with
a standard deviation of 0.44 km.  The non-variability suggests that the
emitting radius area is the complete surface of the NS and not a {\em variable
  fraction}. However, the radius is small with respect to the expected value
\citep[e.g.,][]{Haensel01}. Some of this may be explained by inverse Compton
scattering in a hot electron cloud \citep[see][]{Ebisuzaki87}. Also, we
suspect that the distance estimate of 4.5 kpc may be too low (see discussion).
The other parameters do show variability, but only the variation in peak flux
is (anti) correlated with the state of the persistent flux: when the source is
in the low state it shows a larger dynamical range.  The distribution of the
peak fluxes in the low state is more or less Gaussian with a mean of
$2.85\,\mathrm{cts\,cm^{-2}\,s^{-1}}$ and a variance of $0.35
\,\mathrm{cts\,cm^{-2}\,s^{-1}}$. All the bursts in the high state have peak
fluxes less than $3 \,\mathrm{cts\,cm^{-2}\,s^{-1}}$. The low state
distribution predicts a chance probability of 97.5\% for seeing at least one
burst with a peak flux higher than $3 \,\mathrm{cts\,cm^{-2}\,s^{-1}}$ in the
high state.  Thus, the chance probability of not seeing a burst with a peak
flux higher than $3 \,\mathrm{cts\,cm^{-2}\,s^{-1}}$ in the high state is only
$2.5\%$.  Therefore, bursts during the low state convincingly reach higher
(i.e., up to two times) peak fluxes than during the high state.  We can not
say anything conclusive about the decay times ($\tau$) other than that for all
bursts the decay times are short (i.e., less than 8 seconds) with a weighted
mean of $3.63\, \pm 0.10\,\mathrm{s}$.  This is similar to the bursts from
GX~3+1 seen with other instruments except for the superburst. As mentioned in
the introduction, the longevity of a burst is indicative of the amount of
hydrogen mixture in the helium fuel for the flash. The above decay times
suggest that we are dealing with pure helium burning. In order to double check
this, we averaged all bursts separately for the low and the high state, see
Fig.~\ref{fig:lc_sum}. Thus, the dynamic range of the flux measurements is
increased by at least a factor of 4. This figure shows that 1) there is no
trace of prolonged burst emission and 2) the profiles are identical within the
statistics provided by the data.  For comparison, we added in
Fig.~\ref{fig:lc_sum} the average WFC-measured profile of all long bursts from
GS~1826-24. This source is a well-documented exhibitor of mixed
hydrogen/helium flashes \citep[e.g., ][]{Bildsten00}. These bursts are clearly
a factor of roughly six longer than those of GX~3+1. We conclude that all
bursts from GX~3+1 are short bursts which are thought to be the product of a
thermonuclear runaway process that is fueled by pure helium.

 
\section{Burst rate}
\begin{figure}
\psfig{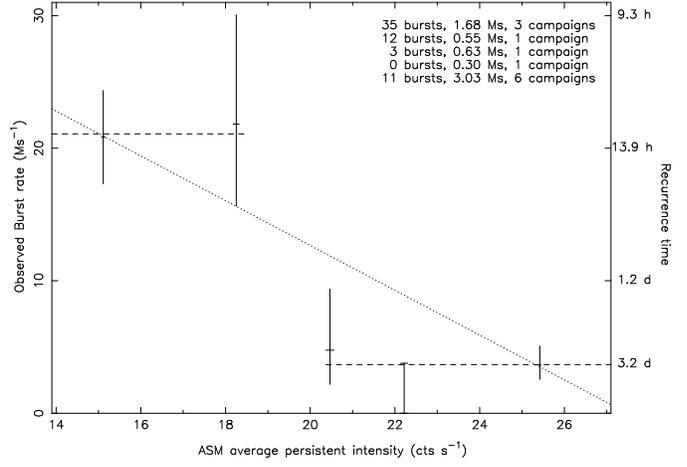}
\caption{Burst rate plotted versus the 1.5--12 keV ASM intensity. The change
  in burst rate is a factor of six. The scale on the right hand side is the
  equivalent recurrence time for the specific burst rate (d for days and h for
  hours). The dashed lines are the best fit of the high- and low burst rate
  level. The dotted line is the best straight line fit.In the upper right
  corner relevant information is given, namely the number of bursts, the net
  observation time and the number of campaigns for each data point from left
  to right.
\label{fig:rate}}
\end{figure}

We have investigated the burst rate as a function of the average persistent
intensity ($\overline{F_{\rm{pers}}}$). To assess the exposure times during
which the sensitivity was sufficient to detect bursts with peak fluxes in
excess of $1\,\rm{cts\,s^{-1}\,cm^{-2}}$ (or 0.5 Crab) we excluded all the
observations whenever the source was located within 30 pixels from the edge of
the FOV. This amounts up to 6\% of the total exposure. For the remaining data
we derived the net exposure times, taking into account Earth occultations and
passages over the South Atlantic Geomagnetic Anomaly (see $t_{\rm{exp}}$ in
Table~\ref{tab:obs}).  The burst rate was calculated by dividing the number of
bursts by the net exposure time for each campaign. The 68\% confidence margins
were determined by assuming Poisson statistics. This results in asymmetric
errors.  As no bursts were seen during the third and the twelfth campaign only
upper limits to the burst rates could be derived. They are $3.8$ and $12.5\,
\rm{Ms^{-1}}$ (68\% confidence), respectively.  For the complete list we refer
to Table~\ref{tab:obs}.  Since the persistent emission and the burst rate of
the different campaigns are sometimes similar, we averaged them over several
campaigns. Campaigns 5, 6 and 7 were averaged for the low state and campaigns
1, 2, 9, 10, 11 and 12 (a null observation) for the high state.  Campaigns 3,
4 and 8 are the campaigns in which the overall intensity decreased or
increased, of which campaign 3 is a null observation. For every campaign we
calculated the average persistent intensity by averaging the ASM measurements
over the complete time span of the campaign. Thus, biases due to the
variability on time scales shorter than the burst recurrence time were
suppressed.  In Fig.  \ref{fig:rate} the burst rate is plotted versus the
persistent emission.  At first sight, one can discriminate two levels of burst
rate, with a quick transition from a high burst rate to a low burst rate at an
intensity between $18.3\,\rm{cts\,s^{-1}}$ and $20.5\,\rm{cts\,s^{-1}}$.  To
test models for the behavior of the burst rate as a function of intensity one
may not use the standard $\chi^2$-minimization, because the statistical
fluctuations are not expected to be distributed as a Gaussian. Instead we
chose to use the statistic

\begin{displaymath}
\label{eq:chi}
S=\sum_{i=1}^{N} \frac{(d_i - m_i)^2}{m_i}
\end{displaymath}

where $d_i$ is the $i$-th measurement, $m_i$ the predicted measurement and $N$
the number of measurements ($N$~=~5). $S$ was minimized through a grid
search. Chance probabilities were evaluated through Monte Carlo simulations,
by counting the number of times that simulated $S$-values were larger than
measured $S$-values.

To find the most probable behavior three models were investigated: a
constant, a linear and a step function. For the constant function no
satisfactorily fit could be found, with a chance probability of
$9\times 10^{-6}$. For the linear model a better fit was found. For a
function of the form $a + bx$ the values $a = 46.46\,\rm{Ms^{-1}}$ and
$b = -1.69\,\rm{Ms^{-1}/cts\, s^{-1}}$ were found. The chance
probability is 21.6\%. The best fit is for the step function. The two
levels that were found are $21.1 \pm 3.0\,\rm{Ms^{-1}}$ and $3.67 \pm
1.0\,\rm{Ms^{-1}}$. The chance probability is 95.3\%. Fig.
\ref{fig:rate} shows the two fits.  If the true relationship is a
linear function, the chance that in a measurement like ours the
minimum $S$ is smaller for a fit with a step function than for one
with a linear function is $7.2\pm0.2\%$, as shown by Monte Carlo
simulations. This is the probability that we are dealing with a linear
function rather than a step function.

\begin{figure}
\psfig{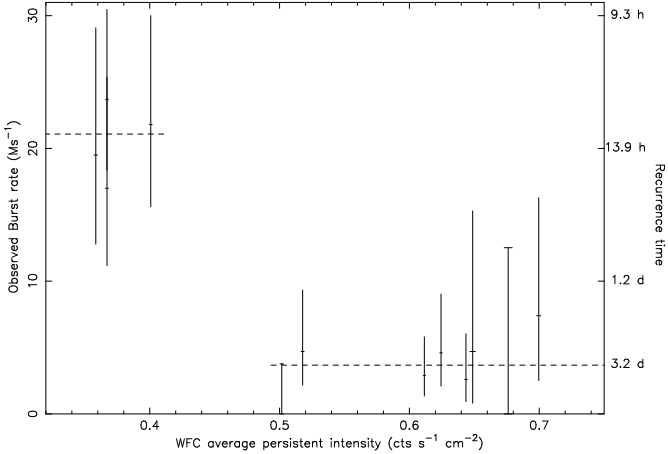}
\caption{Burst rate plotted versus the 2--28 keV WFC flux. In essence is this
  the same plot as Fig.\ref{fig:rate}, but now the persistent flux is averaged
  over smaller time intervals.  Every campaign has been plotted, because the
  spread in persistent intensity is larger. One sees that the transition
  between the high and low burst rate may be broader than in
  Fig.\ref{fig:rate} (i.e., 20\% in stead of 10\%).
\label{fig:rate2}}
\end{figure}

In order to obtain an idea how the transition depends on the definition of the
averaging of the persistent flux, we plot in Fig. \ref{fig:rate2} the burst
rate as a function of the WFC average persistent flux. It shows that the
transition between the high and low burst rate is less narrow. Also the
distribution of the fluxes during the high state is broader. This is the
reason why we have not plotted averages over campaigns like in Fig.
\ref{fig:rate}. The broader transition is also apparent if we plot versus ASM
flux if averaged over the times when the WFCs were on target rather than the
complete time span of each campaign.

\section{Persistent emission and the mass accretion rate}
\label{sec:pers}

At time scales larger than one minute, GX~3+1 exhibits two kinds of
variability \citep{Asai93}. One has a time scale of order minutes to hours and
the modulation depth in the 1--20 keV intensity is typically 20\%.  The
intensity either correlates with the spectral hardness (when the source is
said to be in the 'banana' state) or not ('island' state).  The 7--18/1--7 keV
hardness ratio is observed to change by about 50\%. The other kind has a time
scale of order years and is what causes the source to exhibit the so-called
'low' and 'high' states (Fig.~\ref{fig:asm}). In the following we will
concentrate on the latter (slow) variability because the minimum wait time
between bursts is longer than the time scale of the faster variability; also
the faster variability is presumed to be unrelated to changes in the mass
accretion rate (Asai et al. 1993).  The slow variability has been shown nicely
by 1987--1991 measurements with the all sky monitor on {\em Ginga}
\citep{Asai93}. The trends in those observations are quite similar to the ASM
measurements. The same change in intensity is observed (a factor of 2); such a
variation has been observed in earlier times as well \citep{Makishima83}. The
{\em Ginga}, ASM and earlier data sets even suggest that the flux oscillates
semi-sinusoidally with a period of 6 to 7 years. Interestingly, the {\em
  Ginga} measurements with its Large Area Detector show a constant 1--20 keV
spectral shape over the low and high states \citep{Asai93}.

\begin{figure}[t]
\psfig{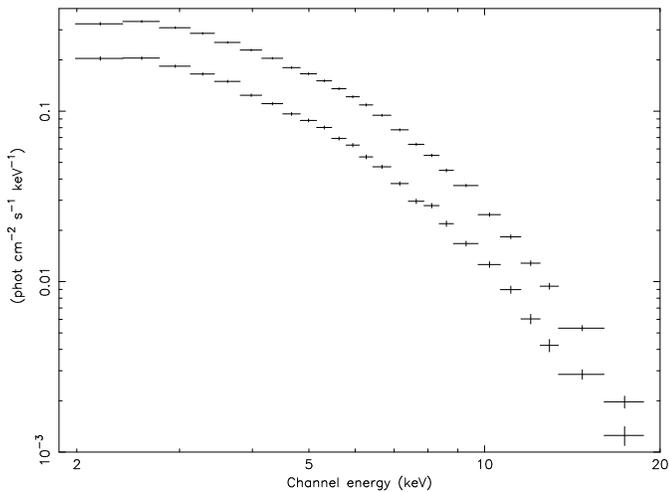}
\caption{Examples of incident photon spectra during low state
  and high state, as measured with the WFC on MJD 51057.24--51058.38
  and MJD 52181.79--52182.83 respectively.
\label{fig:wfcspec}}
\end{figure}

The stability of the X-ray spectrum over the low and high states is more or
less confirmed by the WFC measurements in the 2--28 keV band in the sense that
spectral changes are minimal, see Table~\ref{tabspec} and
Fig.~\ref{fig:wfcspec}. We made a selection of one long observation (i.e.,
longer than one day, except for the last campaign which did not include such a
long observation) per campaign and investigated the full-resolution spectra.
A general fit with an absorbed power law plus disk black body, fixing $N_{\rm
  H}$ again at $1.59\times10^{22}\,\rm{cm^{-2}}$ \citep{ Oosterbroek01} and
coupling the disk black-body temperature and power law index over all 12
spectra, yields $\chi^2/\nu=1.31$ ($\nu=310$; assuming a systematic
uncertainty of 1.5\% and ignoring the insufficiently calibrated spectral
channels 1, 2 and 31 of each spectrum). If the temperature of the disk black
body is allowed to vary over the 12 spectra, the fit improves to
$\chi^2/\nu=1.027$ ($\nu=299$).  Uncoupling the power law index instead of the
temperature yields a slightly worse fit with $\chi^2/\nu=1.06$ ($\nu=299$).
Applying the same model as \citet{Oosterbroek01} on the 0.5--200 keV NFI
spectrum, leaving free only the fluxes of the 3 components, the optical depth
and the black body temperature, but keeping them coupled over the 12 spectra
except for the disk black body normalization, results in a reasonable though
formally not acceptable fit: $\chi^2/\nu=1.354$ ($\nu=320$).  When the disk
black body temperature is allowed to vary, the fit improves to
$\chi^2/\nu=0.945$ ($\nu=309$). Irrespective of the two models, the black body
temperature changes only modestly by about 15\% over the 12 spectra.

\begin{table*}
\caption[]{Results of best-fit spectral models for selection of parameters.
The fits were applied to twelve WFC spectra of GX~3+1, one per campaign.
\label{tabspec}}
\begin{tabular}{lllllllr@{.}l}
\hline

(1)& (2) & (3) & (4) & (5) & (6) & (7) & \multicolumn{2}{l}{(8)}\\
\hline
50319.77--50320.86 & 10.7 & 9.14 & 2.291$\pm$0.013 & 21.0$\pm$0.6 &
2.174$\pm$0.023 & 21.0$\pm$1.0 & 13 & 7\\
50551.19--50552.08 & 10.6 & 9.07 & 2.255$\pm$0.010 & 23.1$\pm$0.6 &
2.140$\pm$0.023 & 22.1$\pm$1.1 & 13 & 6\\
50725.44--50726.54 & 8.95 & 7.74 & 2.234$\pm$0.012 & 20.0$\pm$0.5 &
2.081$\pm$0.027 & 19.9$\pm$1.2 & 11 & 6\\
50910.10--50911.25 & 6.31 & 5.46 & 2.155$\pm$0.019 & 15.2$\pm$0.6 &
1.861$\pm$0.041 & 18.2$\pm$1.8 &  8 & 3\\
51057.24--51058.38 & 6.03 & 5.15 & 2.127$\pm$0.026 & 14.2$\pm$0.8 &
1.768$\pm$0.044 & 20.3$\pm$2.2 &  7 & 8\\
51270.08--51271.21 & 6.60 & 5.62 & 2.211$\pm$0.020 & 13.9$\pm$0.6 &
1.938$\pm$0.041 & 16.2$\pm$1.5 &  8 & 5\\
51422.68--51423.81 & 5.95 & 5.19 & 2.109$\pm$0.020 & 15.9$\pm$0.8 &
1.779$\pm$0.045 & 20.1$\pm$2.2 &  7 & 9\\
51618.05--51619.21 & 8.33 & 7.22 & 2.158$\pm$0.016 & 20.4$\pm$0.7 &
1.966$\pm$0.029 & 22.6$\pm$1.5 & 10 & 9\\
51802.28--51803.42 & 7.98 & 7.87 & 2.256$\pm$0.011 & 17.2$\pm$0.4 &
2.089$\pm$0.032 & 16.4$\pm$1.1 & 10 & 3\\
51989.20--51989.88 & 11.5 & 9.76 & 2.336$\pm$0.022 & 21.2$\pm$1.1 &
2.240$\pm$0.030 & 20.3$\pm$1.2 & 14 & 6\\
52181.79--52182.83 & 11.1 & 9.54 & 2.299$\pm$0.010 & 22.8$\pm$0.5 &
2.203$\pm$0.023 & 21.0$\pm$1.0 & 14 & 3\\
52353.24--52353.71 & 9.72 & 8.34 & 2.293$\pm$0.016 & 20.1$\pm$0.8 &
2.179$\pm$0.029 & 18.3$\pm$1.1 & 12 & 5\\
\hline
\end{tabular}

\noindent
(1) Observation time span (MJD);
(2) unabsorbed 2--28 keV flux (in units of $10^{-9}$~\ecs);
(3) unabsorbed 2--10 keV flux (in units of $10^{-9}$~\ecs);
(4) diskbb k$T_{\rm in}$ (keV);
(5) diskbb $R_{\rm in}^2 cos(\theta)$ (km$^2$);
(6) and (7) as (4) and (5) except in combination with a Comptonized spectrum 
instead of a power law;
(8) unabsorbed 0.01-100 keV flux predicted from the latter model (in units of
$10^{-9}$~\ecs);
\end{table*}

We conclude that the one-day averaged spectral shape shows only a very modest change
over a time scale of years. This change is completely attributable to the
disk black body component. This black body component is likely due to emission
from the inner edge of the accretion disk and is probably driven
by changes in the mass accretion rate \citep[e.g.,][]{Barret01}.

A parameter that is more meaningful than the 2--12 keV photon count rate as
measured with the ASM is the unabsorbed bolometric flux $F_{\rm bol}$ because
that should approximate the mass accretion rate more accurately. If the
spectral model is extrapolated to beyond the WFC bandpass, $F_{\rm bol}$ is
estimated to vary between 8$\times10^{-9}$ and 14$\times10^{-9}$~\ecs\ (see
Table~\ref{tabspec}). Such an extrapolation is justified because a study of
the 0.5--200 keV spectrum by \citet{Oosterbroek01}, taken on MJD 51420-21,
showed that 99\% of the 1--200 keV flux is contained within 1--20 keV.  For a
distance of 4.5~kpc the flux translates to a bolometric luminosity of
2--3$\times10^{37}$~\lum\ (7-12$\times10^{37}$~\lum\ if at a canonical
distance of 8.5 kpc, equal to that of the galactic center).  The smallest
observed flux (8$\times10^{-9}$~\ecs) in combination with the smallest
possible distance (4.5~kpc) and the largest possible NS radius (10~km), while
assuming a NS mass of 1.4~M$_\odot$ and a luminosity given by
$G\,M\,\dot{M}/R$, yields the minimum mass accretion rate in GX~3+1:
1.7$\times10^{-9}$~M$_\odot$~yr$^{-1}$ (or 10\% of Eddington for a
hydrogen-dominated NS atmosphere). It is likely that part of the liberated
gravitational energy is released in another form than radiation.  This
emphasizes that our lower limit for the accretion rate is a firm one. We note
that the mass accretion {\em per unit area} $\dot{m}$ can be estimated
independently from the distance by ($F_{\rm bol}R/GM)(d/R_{\rm app})^2$)
\citep{Bildsten00}. For the neutron star canonical values $\dot{m}$ is above
$2\times10^4$~g~cm$^{-2}$s$^{-1}$.

\noindent


\section{Discussion}
\label{sec:disc}

Our observations of the burst properties as a function of persistent
flux can be summarized as follows. Within a narrow range of persistent
flux of about 10\% (i.e., 18.3 to 20.5 ASM $\rm{cts\,s^{-1}}$), the
average burst frequency drops by a factor of 6 from once per
13.2~hours to once per 3.2~days. Though the best fit is a step
function, we can not rule out a gradual transition. This is the first
time that GX~3+1 is seen to be burst active in the high intensity
state \citep[c.f.,][]{Makishima83, Asai93}. From all other burst
properties only the peak flux seems to be (anti) correlated with the
persistent flux. It occasionally shows higher values, by a factor of
up to 2, during the low state.  In terms of the decay time, there is
no difference between the low and high state (i.e., they are short)
and all bursts appear to be predominantly fueled by helium.  As noted
in the introduction, an anti correlation between persistent flux and
burst frequency is not unexpected.  However, the issues that set
GX~3+1 apart are the quickness of the transition and the fact that on
both sides of the transition only helium bursts are seen.  There is
perhaps one other source, that we know of, that may show a similar
effect but the number of bursts seen in the high state is only two
\citep[4U~1636-53;][]{Lewin87}. The origin of the anti correlation is
not well understood. The fact that in GX 3+1 the anti correlation
takes place over a rather small range of persistent flux may perhaps
be an important clue.

The discrete-like nature of the transition in burst rate is reminiscent of the
boundaries predicted by theory between regimes of unstable nuclear burning on
accreting NSs \citep{Fujimoto81}. For mass accretion rates $\dot{M} >2 \times
10^{-10}$~M$_\odot$yr$^{-1}$, the accreted hydrogen-rich material is expected
to undergo stable hydrogen burning. At the bottom, a helium layer builds up in
which the pressure increases because of the continuing accretion. At a certain
moment the density becomes high enough to ignite the helium in an unstable
burning process giving rise to a thermonuclear flash. If
$\dot{M}<(4-11)\times10^{-10}\,\rm{M_\odot\,yr^{-1}}$, the pressure is high
enough for helium ignition only in the helium layer; however for higher
$\dot{M}$'s, the bottom of the hydrogen layer may also have sufficient
pressure.  As a result, the flash takes place in a mixture of helium and
hydrogen and proton captures occur by the nuclear products of the helium
burning which makes the flash extended because of the longer time scale of the
resulting $\beta$ decays \citep{Taam79}.  Below
$2\times10^{-10}\,\rm{M_\odot\,yr^{-1}}$, the accreted hydrogen no longer
undergoes stable burning, but unstable burning may occur at lower temperatures
because of pressure build-up by the accumulating matter.  The heating produced
by this runaway process may ignite the helium accreted from the companion,
provided the mass of the accumulated layer is large enough. In general, the
amount of hydrogen participating in the flash is larger in this regime than in
the regime $\dot{M}>2\times10^{-10}\,\rm{M_\odot\,yr^{-1}}$.  Finally, there
is a regime above roughly $10^{-8}\,\rm{M_\odot\,yr^{-1}}$ where even the
helium is burned in a stable manner. This is near the Eddington limit and may
not be observable.

The above regimes can experimentally be confirmed through measurements of
burst durations. All observable bursts involve helium burning. For pure
helium, the burst durations are predicted to be of order 10 sec, while the
rise times are shorter than two seconds.  Whenever hydrogen participates, the
burst duration becomes longer. \citet{Paradijs88} confirmed the correspondence
between theory and measurement, although for a limited data set. A
particularly nice illustration of the regimes in a single source is 4U~1705-44
\citep{Langmeier87}.

In GX~3+1, the minimum $\dot{M}$ that is consistent with the persistent flux
is $1.7\times10^{-9}$~M$_\odot\,{\rm yr^{-1}}$. This is near or slightly above
the upper boundary of the range where pure helium bursts are expected to
occur.  Therefore, the data appear discrepant with theory, even if abundances
are sub-solar. Still, the factor-of-two dynamic range of the persistent flux
in GX~3+1 fits comfortably in the factor-of-five dynamic range predicted by
theory for helium bursts. How do other bursters behave that have similar or
higher persistent luminosities? Only a minority of bright LMXBs burst; we can
think of the following four cases. Cyg X-2 is a Z-source with a luminosity
close to Eddington. Thus far, 18 bursts were detected \citep{Kahn84,
  Kuulkers95, Smale98, Titarchuk02}. All are short, like in GX~3+1. The only
other Z-source that exhibits bursts is GX~17+2 \citep[e.g.,][]{Kuulkers02a}.
In that case seven out of ten bursts are clear examples involving hydrogen
burning. Other sources are Cir~X-1 \citep{Tennant86a, Tennant86b} and GX~13+1
\citep{Matsuba95}. While Cir~X-1 shows somewhat longer bursts, the two bursts
detected from GX~13+1 are again short. We propose that the regime of mixed
hydrogen/helium bursts at high accretion rates is seldom observed, that the
regime with pure helium bursts may extend up to higher mass accretion rates,
and that the transition of burst rate in GX~3+1 is not due to change of
regime.

The usual explanation for the anti-correlation is that a larger part
of the helium is burned in a stable manner in the high state
\citep{Makishima83, Paradijs88}. Thus, less fuel is
burnt in thermonuclear flashes. This is supported by the measurement
of the burst parameter $\alpha$ \citep[e.g.,][]{Lewin93} which is the
ratio of the fluence of the persistent emission from burst to burst
and that of the burst. For the low state the average value is
$1.7\times10^3$ and for the high state $2.1\times10^4$. Both these
values are much higher than what is expected and is normal for most
X-ray bursters (between 25 and 250). It is a strong indication that
stable helium and/or hydrogen fusion is happening in both the low and
the high state \citep{Paradijs88}. From the ratio in $\alpha$
the suggestion is that the energy production of the stable fusion
which goes into radiation increases 12-fold from the low to the high
state. Stable helium burning is predicted to be possible only at
near-Eddington accretion rates or above, while GX~3+1 accretes at most
at 30\% of Eddington (see below). However, this difficulty could be
circumvented by assuming that accretion takes place on a confined area
of the NS \citep{Fushiki87a, Bildsten00}. Anyway, if stable helium
fusion is the process responsible for the anti-correlation, our
detection of the rather discrete nature of this anti-correlation
should provide an important test to theory.

There are the open issues of the higher peak fluxes during the low state, the
comparatively small distance for a source that has such a small angular
separation from the galactic center, and the doubtful small radii for the
burst emission areas. We propose that they may all be resolved in the
following manner. \citet{Ebisuzaki83} determined that radiation from helium
bursts may be intense enough to expel the hydrogen atmosphere, thereby leaving
a hydrogen-poor photosphere. Such a photosphere has a higher Eddington limit
so that luminosities may reach higher values by roughly a factor of two. This
is more likely to happen if the hydrogen-dominated atmosphere is thinner. This
is precisely what applies during the low state: the accretion rate is twice as
small and the bursts recur six times quicker so that the accumulated mass from
the companion between bursts is roughly twelve times smaller. Thus, this
provides an explanation why peak fluxes are higher during the low state. The
same explanation has recently been applied to the bimodal peak flux
distribution of bursts from GX 354-0 by \citet{Galloway02}. Unfortunately, in
our case we are not able to confirm the radius-expansion phase in the brighter
bursts due to insufficient statistics.

However, the only radius-expansion burst from GX~3+1 detected by
\citet{Kuulkers00} has a peak flux of 2.3 Crab and happened during the low
state.  The peak flux compares well, within a few tens of percent, with the
maximum peak flux that we find during the low state. The above reasoning
implies that this burst reaches the Eddington limit for a helium-rich
photosphere. \citet{Kuulkers00} determined that in that case (X=0) the
distance would be $6.1\pm0.1$~kpc. This, in comparison to the 4.5~kpc, is
substantially closer toward the galactic center (at 8.5~kpc) and seems more
plausible. Furthermore, the radius of the burst emission scales up accordingly
to a more plausible value of $6.05\pm0.11$~km (weighted mean). The luminosity
of the persistent emission would increase accordingly. The maximum luminosity
would be roughly $6\times10^{37}$~\lum, or 30\% of Eddington. We note that
6.05 km is still smaller than the expected NS radius \citep{Haensel01}. This
may be explained by inverse Compton scattering effects as discussed by
\citet{Ebisuzaki87} which may introduce a factor of 2 decrease in the apparent
radius for the black-body temperatures seen in GX~3+1.

\begin{acknowledgements}
  We would like to thank colleagues at the {\em BeppoSAX} Science Operations
  Center for their devoted support and the ASM/RXTE team for the repair of
  the light curve of GX~3+1. JZ and EK acknowledge
  financial support from the Netherlands Organization for Scientific Research
  (NWO).
\end{acknowledgements}

\bibliography{literature}

\end{document}